# A Community-Developed Open-Source Computational Ecosystem for Big Neuro Data

Randal Burns, Eric Perlman, Alex Baden, William Gray Roncal, Ben Falk, Vikram Chandrashekhar, Forrest Collman, Sharmishtaa Seshamani, Jesse Patsolic, Kunal Lillaney, Michael Kazhdan, Robert Hider Jr., Derek Pryor, Jordan Matelsky, Timothy Gion, Priya Manavalan, Brock Wester, Mark Chevillet, Eric T. Trautman, Khaled Khairy, Eric Bridgeford, Dean M. Kleissas, Daniel J. Tward, Ailey K. Crow, Matthew A. Wright, Michael I. Miller, Stephen J Smith, R. Jacob Vogelstein, Karl Deisseroth, and Joshua T. Vogelstein

**Big imaging data is becoming more prominent in brain sciences across spatiotemporal scales and phylogenies. We have developed a computational ecosystem that enables storage, visualization, and analysis of these data in the cloud, thusfar spanning $20+$ publications and $100+$ terabytes including nanoscale ultrastructure, microscale synaptogenetic diversity, and mesoscale whole brain connectivity, making NeuroData the largest and most diverse open repository of brain data.**

Recent developments in technology, such as high-throughput imaging and sequencing, enable experimentalists to collect increasingly large, complex, and heterogeneous "big" data [1]. Any study includes both raw data and metadata, potentially resulting in terabytes per day, eventually yielding petabytes when aggregating across experiments and laboratories. These new experimental capabilities exceed the scale or feature-set of existing software. For example, such data cannot be stored, processed, and visualized on a laptop or workstation. Instead, big data require complex registration, processing, and machine learning pipelines, to be stored on data centers and processed on high-performance and/or cluster computers.

There is a natural inclination therefore to begin building custom, bespoke software solutions. Indeed, many laboratories around the world have their own software stack and practices, maintained by ever shifting graduate students or software engineers, using local hardware resources that must also be maintained. This approach is difficult to translate to future experiments, individuals, or institutions. In NeuroData's first efforts in the Open Connectome Project (http://neurodata.io), we developed a complex software stack to manage terascale spatial data for high-throughput electron microscopy [2]. This stack hosted our collaborators' data from 2011 to 2015, but our academic development team became overwhelmed as technology changed, features were added, and scale increased.

We therefore developed an open-source, community-built, software ecosystem deployed in the commercial cloud. This required integrating multiple open-source projects and extending them for our specific needs. As new scientific questions demand new features, we work with open-source developers spanning organizations and institutions, incorporating features into their tools or adding "plug-ins" that we maintain (Figure 1).[1] As new features were required to address our scientific questions, we worked with the primary developers spanning organizations and institutions, and either incorporated our features into their tools, or added "plug-ins" that we maintain.

Data visualization helps researchers evaluate sample preparation, quality assurance, image processing, and annotation, and facilitates hypothesis generation and discovery [3]. Thus, NeuroData prioritizes visualization via a three-dimensional (3D) interactive Web application that connects to all data sources, including raw image tiles off the microscope, stages of transformation, and published annotated stacks. Users zoom, rotate, and pan among multiple channels, canonical planes, and rendered volumes. Scientists interact with volumetric structures to understand synapse morphology or sparse protein expression. Several use cases illustrate how visualization enables scientific discovery.

Serial electron microscopy (EM) produces hundreds of terabytes of monochromatic image tiles [4].

Their life-cycle includes storing raw images, stitching them into 3D volumes, correcting intensity aberrations, and extracting objects of interest, including neural processes and synapses. Fiji plugins and other tools solve the stitching and registration problem [5].[2] We built and modified tools for discovery on registered and aligned data (Fig. 2, top left). For visualization, we built NeuroDataViz (NDViz) by forking and extending Google's NeuroGlancer, adding multi-channel overlays, data sources, and minor extensions to the WebGL core. We modified "distributed multi-grid" (DMG)—originally developed for natural images and cosmology—to address intensity biases across images [6]. Finally, many neuroimaging voxels are 10x anisotropic. Rotating 3D annotations in space therefore produces a stair-step look, which we mitigated by co-opting a mesh smoothing tool [7].

Conjugate array tomography identifies 20+ channels of protein expression in 3D, followed by serial electron microscopy [11]. The life-cycle of these data includes storing raw image tiles, registering and stitching each channel into 3D volumes, co-registering channels and modalities, and extracting objects of interest. For data visualization, we extended Render into the cloud. Render stores and applies image transforms that describe how volumes are assembled. Render does not provide algorithms, but stores their output to reproduce all pipeline stages on demand. Array tomography channels characterize protein expression co-registered to electron microscopy structure. Underlying data are stored as raw image tiles. The top right image in Figure 2 is constructed on demand, which minimizes computation and data movement. Scientists can annotate crude alignments, labeling neurons or synapses. Render carries these annotations forward into improved future alignments. For this project, we extended NDViz and NeuroGlancer to support multi-spectral images, using Render as a data source.

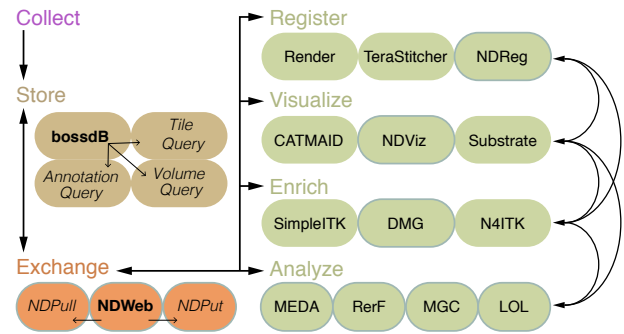

Figure 1: Neurodata's open-source software ecosystem. Software projects that NeuroData develops and maintains have a gray border. Note that even those projects heavily leverage open source packages. For example, NDViz is a fork of Google's Neuroglancer, and NDReg uses SimpleITK for linear and affine pre-registration.

CLARITY is a method of making brain tissue transparent to enable estimating the brain-wide joint statistics of multiple neurons [12]. CLARITY experiments collect data from multiple channels, including background for registration, and a "signal" channel that exhibits protein expression localized to a particular sub-population of neural cells. CLARITY Optimized Light-sheet Microscopy (COLM) can accelerate data collection from clarified samples by several orders of magnitude while maintaining or increasing quality and resolution, yielding multi-terabyte datasets [13]. After data are stitched together with TeraStitcher [14], the volume is registered to an atlas to identify relative expression across brain regions [15]. We built an open-source package, NDReg, implementing multiscale large deformation diffeomorphic metric mapping, a principled mathematical theory of nonlinear registration devised for computational anatomy of magnetic resonance imaging data [16, 17]. This tool registers these multi-terabyte images in under an hour on a workstation. NeuroData also built Web-accessible microservices to enrich visualization. Our open-source service provides atlas metadata, allowing visualization tools to dynamically display brain regions associated with images (Fig. 2, bottom).

At publication, image data enter an archival database that provides high-throughput access to volumetric data. Spatial databases serve multiple purposes. (1) They enable visualization over the Internet, updating the screen at tens of frames per second. (2) They download resolutions and regions on de-mand. (3) They provide data to analysis algorithms, such as computer vision that identifies neurons, axons, and dendrites. NeuroData deploys the bossDB, developed for EM and co-registered functional image and annotation data [18]. Based on the Open Connectome Project, bossDB provides spatial databases as a service in the cloud. This software was developed at the JHU Applied Physics Laboratory under a partnership with the IARPA MICrONS program, and exemplifies avenues for growth and support toward enabling community research.

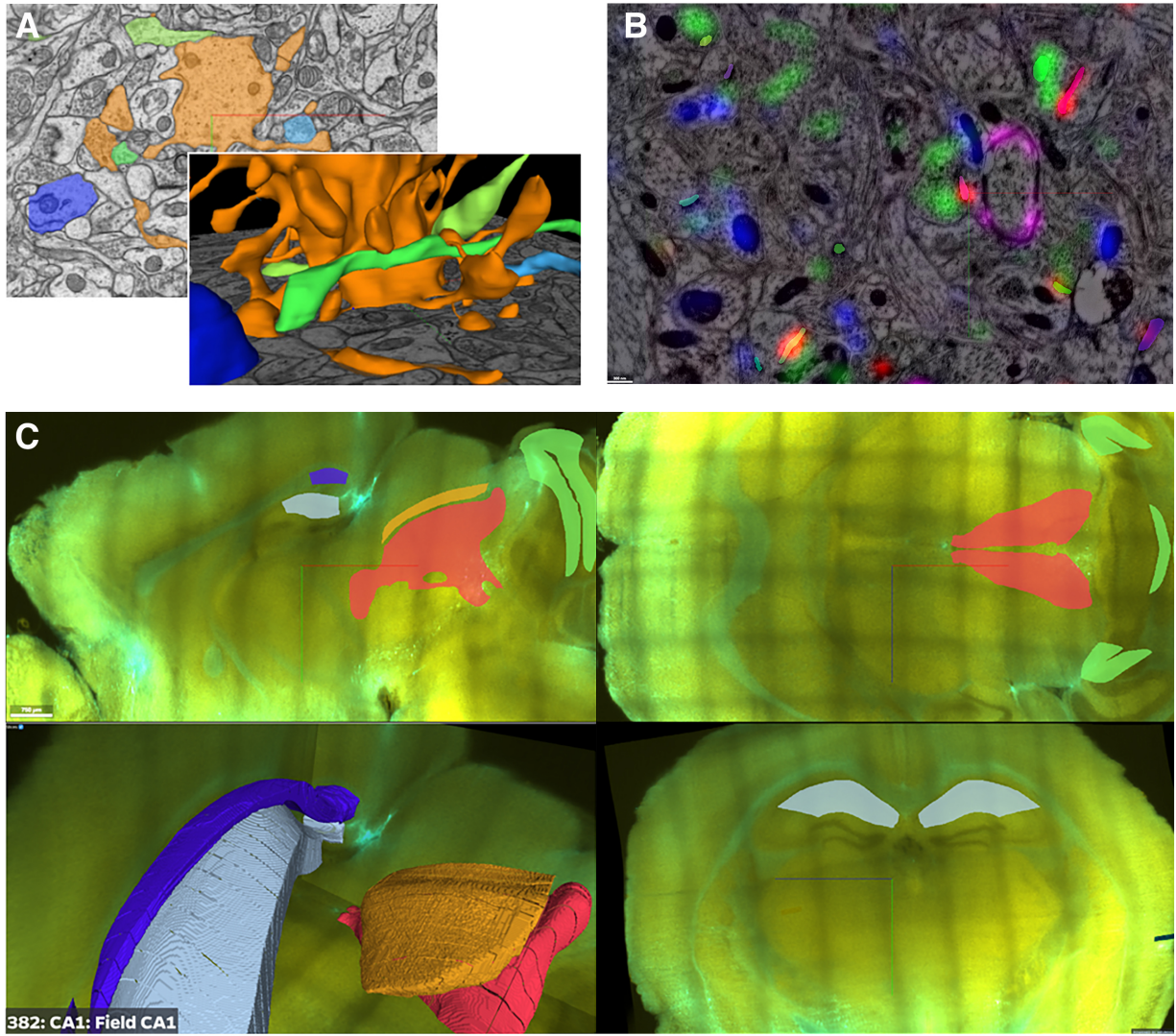

Figure 2: **Interactively visualizing brain volumes.** **(A)** 3D volume of manual annotations intersecting a plane of EM data [8] displayed over 2-dimensional image acquisition plane. **(B)** Multi-channel view comparing co-registration of EM with multi-spectral light-microscopy protein expression [9]. Color blending facilitates visualizing multiple colors. Data accessed from Render. **(C)** CLARITY [10] coregistered to Allen reference atlas and linked with metadata microservice that displays brain region names under cursor. Image data accessed from bossDB and visualized using NDViz, after registration with NDReg. The cloud database, visualization, registration, and metadata let users navigate images by brain region.

NeuroData holds 100 public and private datasets, 200 teravoxels from 30 collaborators, making it the world's largest and most diverse public neuroscience data repository. We host approximately 140 teravoxels of public image data, from nanometer to millimeter scales (Table 1). Registered to these databases, annotation datasets provide semantic objects for volumetric analysis. Thus, scientists can investigate image quality from different technologies before deciding which ones to adopt. Moreover, scientists can access and analyze disparate data with the same functionality and syntax, to develop more coherent and comprehensive models of the function and dysfunction of nervous systems.

Table 1: Image Datasets in Open Connectome Project. Modalities: EM = Electron Microscopy, AT = Array Tomography, Ophys = Optical Physiology, XCT = X-ray Micro Computational Tomography, LM = Light Microscopy, CL = CLARITY, MR = Magnetic Resonance. Bits = number of bits representing image intensity per voxel. Proj = number of databases in dataset. Ch = channels across dataset. T = timesteps across dataset. GV = approximate number of gigavoxels ($10^9$ voxels) across project. Total, 139,874 gigavoxels across 23 publications with 52 projects containing 527 channels. For references, licenses, and DOIs, https://neurodata.io.

| Reference | Modality | Species | Bits | Proj | Ch | T | GV |
|---|---|---|---|---|---|---|---|
| Bhatla | EM | C. elegans | 8 | 1 | 2 | 1 | 226 |
| Jarrell | EM | C. elegans | 8 | 1 | 1 | 1 | 43 |
| White | EM | C. elegans | 8 | 1 | 1 | 1 | 28 |
| Bumbarger | EM | P. pacificus | 8 | 1 | 1 | 1 | 1672 |
| Hildebrand | EM | D. rerio | 8 | 1 | 1 | 1 | 26,869 |
| Wanner | EM | D. rerio | 8 | 1 | 1 | 1 | 1,108 |
| Tobin | EM | D. melanogaster | 8 | 1 | 1 | 1 | 55,113 |
| Ohyama | EM | D. melanogaster | 8 | 2 | 1 | 1 | 2,609 |
| Takemura | EM | D. melanogaster | 8 | 1 | 1 | 1 | 210 |
| Bock | EM | M. musculus | 8 | 1 | 1 | 1 | 20,102 |
| Lee | EM | M. musculus | 8 | 1 | 1 | 1 | 22,065 |
| Kasthuri | EM | M. musculus | 8 | 3 | 1 | 1 | 1,544 |
| Harris | EM | R. rattus | 8 | 3 | 3 | 1 | 19 |
| Collman | EM&AT | M. musculus | 8 | 2 | 26 | 1 | 46 |
| Bloss | AT | M. musculus | 8 | 1 | 3 | 1 | 493 |
| Weiler | AT | M. musculus | 16 | 12 | 288 | 1 | 215 |
| Micheva | AT | M. musculus | 16 | 1 | 26 | 1 | 25 |
| Dyer | XCT | M. musculus | 8 | 1 | 1 | 1 | 3 |
| Kutten | CLARITY | M. musculus | 16 | 12 | 23 | 1 | 7,191 |
| Branch | iDISCO | M. musculus | 16 | 1 | 1 | 1 | 5 |
| Vladimirov | Ophys | D. rerio | 16 | 1 | 1 | 100 | 9 |
| Randlett | LM | D. rerio | 16 | 1 | 138 | 1 | 16 |
| Amunts | MR | H. sapiens | 8 | 1 | 1 | 1 | 262 |
| Grabner | MR | H. sapiens | 16 | 1 | 3 | 1 | <1 |
| Totals | – | – | – | 52 | 527 | – | 139,874 |

**Acknowledgments** RB, EP, AB, BF, VC, KL, MK, EB, DJT, MIM, and JTV are at Johns Hopkins University, supported by the Defense Advanced Research Projects Agency (DARPA) SIMPLEX program, SPAWAR contract N66001-15-C-4041, APL Internal Research and Development Funds, NIH-TRA 1R01NS092474, "Synaptomes of Mouse and Man," and NSF 16-569 NeuroNex. RJV and DMK are at Gigantum. EMT and KK are at Janelia Research Campus. MC is at Facebook. FC, SS, and SJM are at the Allen Institute for Brain Sciences. AKC, MAW, and KD are at Stanford University. WGR, BW, RH, TG, PM, and JM are at the Johns Hopkins University Applied Physics Laboratory. Their work supported by the Office of the Director of National Intelligence (ODNI), Intelligence Advanced Research Projects Activity (IARPA), via IARPA Contract No. 2017-17032700004-005 under the MICrONS program. The views and conclusions contained herein are those of the authors and should not be interpreted as necessarily representing the official policies or endorsements, either expressed or implied, of the ODNI, IARPA, or the U.S. Government. The U.S. Government is authorized to reproduce and distribute reprints for Governmental purposes notwithstanding any copyright annotation.

## Notes

[1] Including https://github.com/jhuapl-boss, https://github.com/neurodata/ndpush, https://github.com/neurodata/ndwebtools, https://github.com/neurodata/ndpull, https://github.com/saalfeldlab/render, https://opensource.google.com/projects/neuroglancer, http://abria.github.io/TeraStitcher/, https://github.com/neurodata/ndreg, http://catmaid.org, https://github.com/neurodata/ndviz, https://github.com/iscoe/substrate, http://www.simpleitk.org/ https://github.com/neurodata/ndmesh, https://github.com/mkazhdan/DMG, https://github.com/neurodata/R-RerF, https://github.com/neurodata/MGC, and https://github.com/neurodata/LOL.

[2] https://github.com/khaledkhairy/EM_aligner and https://github.com/billkarsh/Alignment_Projects